# GB1508+5714: a radio-loud quasar with z=4.30 and the space density of high redshift, radio–loud quasars

I. M. Hook[1] *, R. G. McMahon[1], A. R. Patnaik[2] †, I. W. A. Browne[2],
P. N. Wilkinson[2], M. J. Irwin[3], C. Hazard[1,4]

[1] *Institute of Astronomy, Madingley Road, Cambridge CB3 0HA, UK*
[2] *University of Manchester, Nuffield Radio Astronomy Laboratories, Jodrell Bank, Macclesfield, Cheshire SK11 9DL*
[3] *Royal Greenwich Observatory, Madingley Road, Cambridge, CB3 0EZ*
[4] *University of Pittsburgh, Pittsburgh, PA 15260, USA*

**ABSTRACT**

We report the discovery of a radio loud quasar with a redshift of 4.30. This object, which is the first radio selected quasar with a redshift greater than four, was discovered during an observational investigation into the evolution of the luminosity function of radio loud quasars. Here we describe results based on a sample of ∼300, flat spectrum radio sources with $S_{5GHz}$>200mJy. In this study, rather than carry out an indiscriminate redshift campaign on all the radio sources, we have used the APM POSS catalogue to preselect a subset of stellar-like optical counterparts with red optical colours. Such a subsample is expected to contain a high fraction of high redshift quasars. 10 of the ∼300 sources were selected for follow-up optical spectroscopy and three of these are identified as quasars with redshifts greater than 3.0. One of the radio sources, GB1508+5714, is a quasar with $z$=4.30 and m(R)∼19.

## 1 INTRODUCTION

Quasars, by virtue of their extremely high intrinsic luminosities over most of the observable electromagnetic spectrum, should be detectable out to redshifts of more than 10. Thus observations of high redshift quasars are powerful tools in observational cosmology.

Whilst the first quasars were identified via their radio emission, the majority of quasars are currently discovered using optical survey techniques. This is primarily because the space density of radio-loud quasars is much lower than that of optically-selected quasars. In the redshift range 2<z<4.5 the fraction of optically-selected quasars that are radio-loud at even the mJy level is only ∼10% (McMahon 1991). However, it is possible to carry out surveys for radio-loud quasars over very large areas of sky, and thereby still obtain sufficient numbers of sources for a statistical analysis of their space density.

Radio selection is potentially more powerful than optical techniques since radio-selected QSO samples can be defined using simpler selection criteria, and their radio fluxes are unaffected by any obscuration due to dust in foreground galaxies. The spectral energy distribution of quasars at radio wavelengths is smooth and is not complicated by either emission lines or redshift-dependent intervening absorption so that the effects of k-corrections are easier to model. Moreover, since the mean spectral index of core-dominated, flat-spectrum radio sources is around 0.0, a higher fraction will lie at high redshift compared with optical or X-ray samples which have typical spectral indices in the range $-0.5$ to $-1.0$.

In this paper we describe initial results of a survey for radio-loud quasars with redshifts greater than 3. The survey exploits the observation that quasars at high redshift have redder optical colours than their low-redshift counterparts due to absorption by intervening Lyman-α. Throughout this paper we have assumed cosmological constants of $H_0 = 50\,\text{km}\,\text{s}^{-1}\,\text{Mpc}^{-1}$ and $q_0 = 0.5$.

## 2 THE RADIO SAMPLE AND THE OPTICAL IDENTIFICATION PROCEDURE

The sample used in this study is a subset of the ∼ 1600 flat-spectrum ($\alpha_{1.4GHz}^{5GHz} \geq -0.5$, $S \propto \nu^\alpha$) radio sources with $S_{5GHz} > 0.2Jy$ studied with the VLA by Patnaik et al.

\* Present address: UC Berkeley Astronomy Dept., Berkeley CA94720, USA.
† Present address: Max-Planck-Institut fuer Radioastronomie, Auf dem Hugel 69, D-53121 Bonn, Germany

(1992). This sample was chosen for the present study for its relatively large area, suitable flux limit, high frequency (5GHz) and content of flat-spectrum sources.

The Automated Photographic Measuring (APM) facility at Cambridge was used to measure all the blue (O) and red (E) POSS (Palomar Observatory Sky Survey) plates within the limits $7^h < \alpha(J2000) < 19^h$, $35° \leq \delta(J2000) \leq 75°$ and $|b| \geq 30°$ (1.02sr). About 300 sources from an area of 0.72sr within the above region were identified prior to the scheduled observing run in April 1992.

Optical identifications were made on the basis of positional coincidence alone. The *rms* difference between optical and radio positions, $\Delta r$, was found to be $0.7''$ and the criterion for positional coincidence was defined as $\Delta r \leq 3.0''$.

APM image classification software was used to classify each object found on the E (reference) plate as stellar or extended (*ie* galaxy/merged). The O and E catalogs were then matched in position and a colour (O−E) was calculated for each object that appeared on both plates. Colour limits of objects only detected on the E plate were also calculated. This is important since many red objects are only detected on the E plate.

The zero point for photometry in each field was calculated by assuming the E plate has a limiting magnitude of 20.0 and by assuming a universal, magnitude independent position in the colour-magnitude plane for the stellar locus, as described in McMahon (1991). The zero point of the magnitude system has an rms uncertainty of 0.25, derived from the comparison of objects detected in the overlapping regions of plates and CCD photometry in 11 of the fields used in the identification procedure.

From the catalogue of optical sources within $3.0''$ of a radio source and with $E \leq 19$ we then constructed a list of candidate high-redshift quasars for spectroscopic follow-up. To be considered, an optical counterpart had to be classified as stellar on the E plate and had to have a red O−E colour. In figure 1 we show the dependence of O−E colour on redshift for a sample of optically-selected QSOs. The sample consists of a selection of QSOs from the catalogue of Hewitt & Burbidge (1989) supplemented by $z > 3$ objective prism selected objects (eg. Hazard, McMahon, & Sargent 1986) and objects from the APM BRI survey with $z > 4$ (Irwin, McMahon & Hazard 1991).

Since we are primarily interested in $z > 3$ QSOs, we imposed a colour cut of O−E>1.5. This colour selection eliminates the majority of the low-redshift quasars (see below). This defines a complete sample of 10 objects, one of which (B2 1048+3446) was a previously-known QSO with $z = 2.52$ (Wampler *et al.* 1984). The remaining 9 objects were observed spectroscopically as described below.

## 3 SPECTROSCOPIC OBSERVATIONS AND RESULTS

Spectroscopic observations were first carried out during the period 20 April to 25 April 1992, with the 4.3m William Herschel Telescope (WHT) and using the Faint Object Spectrograph (FOS) (Allington-Smith *et al.* 1989). FOS is a cross-dispersed low resolution spectrograph and useful data was obtained only in first order (8.7Å/pixel) over the wavelength range 5000–9500Å. Exposure times were short, typi-cally 300secs and the spectra were adequate for initial classification and redshift determination.

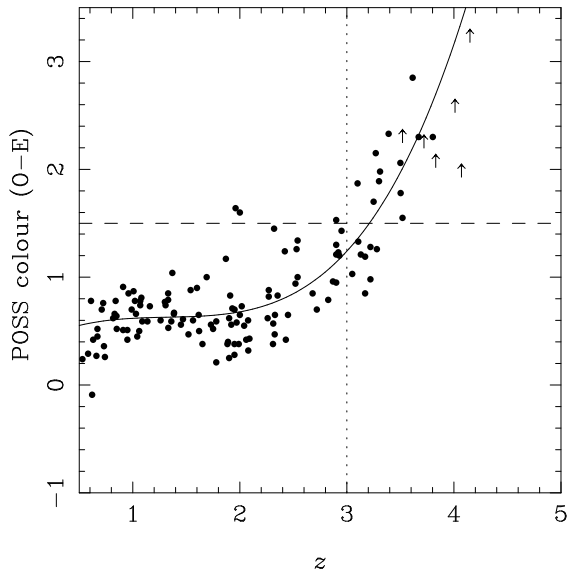

**Figure 1.** O−E vs $z$ for a sample of optically-selected quasars with a polynomial fit superimposed. The horizontal line at O−E=1.5 shows the limit used to define our complete sample. Arrows represent lower limits on O − E for objects detected only on the E plate.

The optical spectra of three of the target objects were characterised by strong, broad emission lines. These objects were identified as high-redshift quasars by the presence of broad Ly$\alpha$ (rest wavelength 1216Å) and CIV (1549Å) and by the drop in flux across the Ly$\alpha$ emission line caused by the Ly$\alpha$ forest. Subsequently, spectra of higher signal-to-noise and at higher resolution were obtained of the two higher-redshift objects, GB1508+5714 and GB1745+6227. These spectra were obtained using ISIS (Intermediate-dispersion spectrograph and Imaging System) on the WHT. On the red arm a 158 l/mm grating and a EEV CCD with 1152x1242 22.5$\mu$m pixels were used and on the blue arm a 158 l/mm grating and a thinned Tektronix with 1024x1024 24$\mu$m pixels.

The quasar GB1508+5714 was re-observed on 17th April 1993 with the WHT and the ISIS double spectrograph. Three 900s observations were obtained with the red arm and two 1500s with the blue arm. The quasar GB1745+6227 was observed on Aug 1 1992 within the context of the PATT service observation program. Seven 900s observations were obtained with the red arm and six 900s with the blue arm. In both cases a $1.5''$ spectrographic slit aligned to the parallactic angle was used. Red and blue observations of the flux standards BD +26° 2606 and BD +40° 4032 were used to obtain relative flux calibration over the observed wavelength range for GB1508+5714 and GB1745+6227 respectively.

The redshifts of the three new quasars are 3.10, 3.89 and 4.30, the latter being the highest redshift radio-selected quasar known. The properties of these quasars are summarised in table 1. The spectra of the three objects are shown in figure 2 and the determination of their redshifts are discussed in turn below. A finding chart for GB1508+5714 is given in figure 3.

**Table 1.** Properties of the new $z > 3$ quasars.

| Name | Optical position $\alpha$ (B1950) $\delta$ | $z$ | E mag | O-E mag | $S_{5GHz}$ mJy | Feature | $\lambda_{obs}$ Å | $z_{meas}$ |
|---|---|---|---|---|---|---|---|---|
| GB1508+5714 | 15 08 45.19 +57 14 02.6 | 4.301 ±0.006 | 18.9 | >3.44 | 282 | Ly$\alpha$ | 6451.1 | 4.305 |
|  |  |  |  |  |  | NV | 6566.3 | 4.295 |
|  |  |  |  |  |  | CIV | 8213.1 | 4.302 |
| GB1745+6227 | 17 45 47.98 +62 27 55.2 | 3.889 ±0.001 | 18.3 | 2.61 | 580 | Ly$\alpha$ | 5944.8 | 3.890 |
|  |  |  |  |  |  | SiIV/OIV] | 6833.3 | 3.889 |
|  |  |  |  |  |  | CIV | 7573.7 | 3.889 |
| GB1338+3809 | 13 38 11.09 +38 09 53.5 | 3.103 ±0.002 | 17.9 | 1.71 | 305 | Ly$\alpha$ | 4985.3 | 3.101 |
|  |  |  |  |  |  | SiIV/OIV] | 5737.2 | 3.104 |
|  |  |  |  |  |  | CIV | 6356.7 | 3.104 |

Redshifts were measured as described in the text, assuming rest-frame wavelenghts for the emission lines as follows: Ly$\alpha$:1215.7, CIV: 1549.1, NV:1240.1, SiIV/OIV:1397.8. The radio fluxes are from Gregory and Condon (1991).

**GB1508+5714** Broad emission lines of Ly$\alpha$ (rest wavelength 1215.7Å), NV (1240.1Å) and CIV (1549.1Å) are visible in the spectrum. The CIV line is seen at 8213.1Å, (Gaussian line centroid) giving a redshift of 4.30. The Ly$\alpha$ line is visibly a-symmetric due to absorption by intervening neutral Hydrogen, hence its observed wavelength of 6451.1Å was measured from the peak rather than by determining the centroid of the line. This observed wavelength gives an upper limit to the redshift of 4.31. NV is observed at 6566.3Å, giving a redshift of 4.30. Also visible in the spectrum is the weak SiIV/OIV] blend at ~7400Å. The redshift of this object was taken to be 4.30.

**GB1745+6227** Broad emission lines of Ly$\alpha$, NV and CIV are visible in the spectrum. Again the Ly$\alpha$ line has visible absorption, and the bluer of the two peaks lies at an observed wavelength of 5944.8Å giving an upper limit to the redshift of 3.89. The CIV line has an observed wavelength of 7573.7Å, giving a redshift of 3.89. The SiIV/OIV] blend is also visible at 6833.3Å. Assuming a rest wavelength of 1397.8Å for this line, this also gives a redshift of 3.89.

GB1745+6227 was independently discovered by Becker, Helfand & White (1992) on the basis of its X-ray emission. They report a redshift of 3.87, which is lower than our determination. Stickel (1993) reports a redshift of 3.886.

**GB1338+3809** Ly$\alpha$ and CIV emission lines are clearly visible in the spectrum. The peak of the Ly$\alpha$ line lies at an observed wavelength of 4985.3Å giving an upper limit to the redshift of 3.10. CIV is observed at 6356.7Å and the SiIV/OIV] blend at 5737.2Å, both giving a redshift of 3.10. Assuming a redshift of 3.10, the CIII emission line (rest wavelength 1908.7Å) lies at 7825.7Å. However, this line is extremely weak in this object and is bearly detected above the continuum.

The remaining objects had no obvious emission features in their spectra and are likely to be low-redshift ($z < 0.5$) galaxies or BL Lac objects which are spatially unresolved at the POSS resolution.

In summary, the complete sample of 10 objects defined in section 2 contains 3 high-redshift quasars with redshifts

**Table 2.** Predicted and observed numbers of high-redshift, flat spectrum quasars.

| Model | Number | | | |
|---|---|---|---|---|
|  | 3.0<z<3.5 | 3.5<z<4.0 | 4.0<z<4.5 | Total |
| (a) | 3.8 ± .2 | 4.7 ± .3 | 3.3 ± .2 | 11.7 ± .7 |
| (b) | 1.5 ± .1 | 1.4 ± .1 | 0.70 ± .05 | 3.6 ± .2 |
| (c) | 0.89 ± .05 | 0.70 ± .04 | 0.28 ± .02 | 1.9 ± .1 |
| Obs | 1 | 1 | 1 | 3 |

of 3.10, 3.89 and 4.30, one previously-known $z = 2.52$ quasar and 6 low-redshift objects.

## 4 DISCUSSION

We now compare the observed number of $z > 3$ quasars with model predictions based on the luminosity function derived at lower redshifts. The expected number of flat-spectrum quasars in our sample was computed using the $z = 2$ luminosity functions derived from models $1 - 5$ in Dunlop & Peacock (1990), hereafter DP90. These DP90 models concern the population of all flat-spectrum radio sources which is strongly dominated by quasars. We consider simple density evolution for $z$>2 and have computed the expected number for a constant comoving density and for cases where the space density decreases by a factor of 2 and 3 per unit redshift from the $z = 2$ value (models (a), (b) and (c) in table 2 respectively). The predicted numbers have been calculated for a survey area of 0.72sr and allow for incompleteness due to the colour cut. The *relative* completeness has been calculated by considering the residuals from a polynomial fit to the colour-redshift data in figure 1. The completeness increases from ~ 50% to ~ 100% over the range $3.0 < z \leq 3.5$ and is ~ 100% in the range $3.5 < z < 4.0$. For $z > 4.5$ the completeness of our survey drops rapidly due to absorption of the E passband by the Ly$\alpha$ forest. The mean and range of predictions in the redshift range $3.0 < z < 4.5$ are tabulated in table 2 with the observed numbers. ¿From table 2 it appears that the numbers of quasars found in the range $3.0 < z < 4.5$ are consistent with a model in which the co-

4  *I. M. Hook et al.*

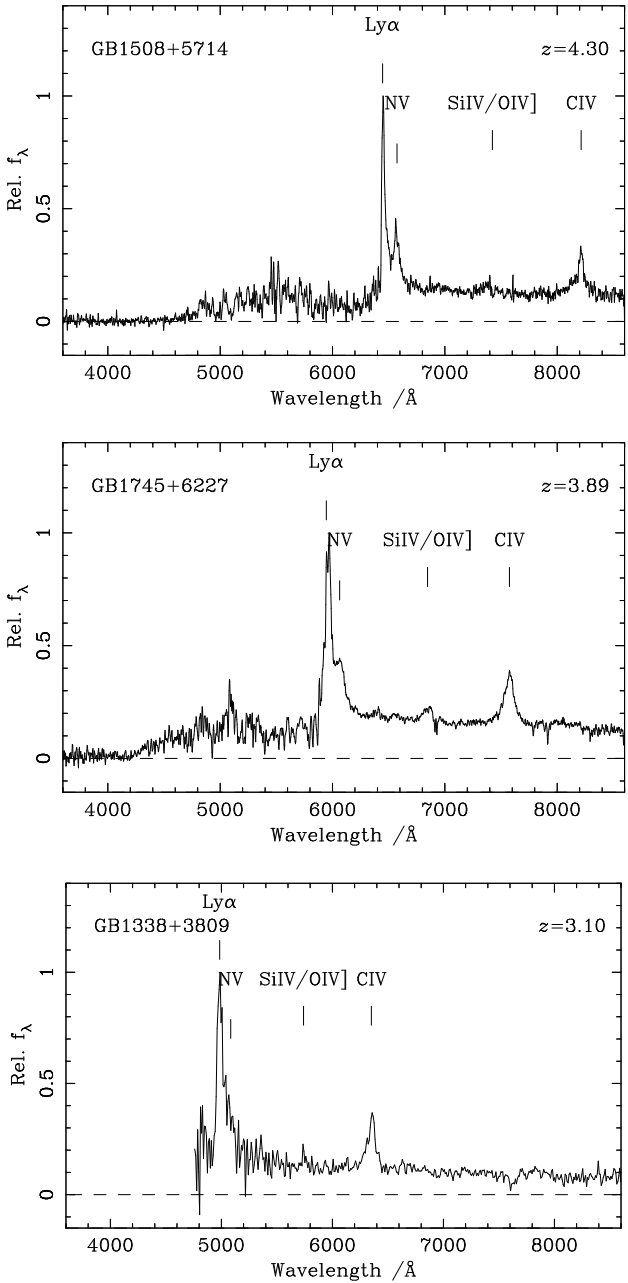

**Figure 2.** Spectra of the three quasars with $z > 3$.

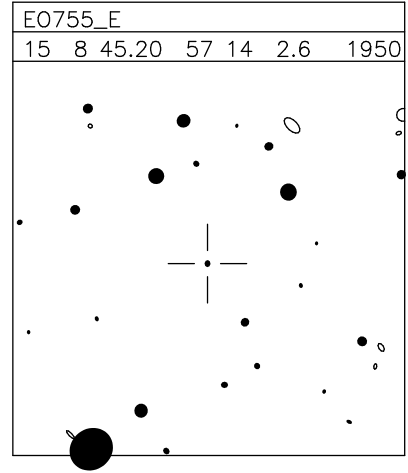

**Figure 3.** A finding chart for GB1508+5714. The central cross is 1 arcmin across. The solid ellipses indicate objects classified at 'stellar' by the APM, and open ellipses represent extended objects. North is at the top, West to the right.

moving density decreases by a factor of 2-3 per unit redshift at redshifts above 2.

## 5 CONCLUSIONS

Whilst the original models of DP90 are well-constrained at $z \sim 2$, there is a large dispersion in the predictions of these models for $z > 3$. Models 1−5 in DP90 predict that between 3 and 7 objects with $3.0 < z < 4.5$ would be found in our survey (allowing for incompleteness due to the colour cut). The results presented here provide tighter constraints on the luminosity function of radio-loud quasars with $z > 3$, the subject of a later paper.

Extrapolating models (b) and (c) in table 2, we would expect there to be between 1 and 5 radio-loud ($S > 200mJy$) quasars between $z = 5$ and $z = 6$ over the whole sky (*ie.* $4\pi$sr). These objects will be among those too faint to be identified on the POSS plates, hence we intend to obtain deep CCD observations of the blank fields.

In the meantime we are extending our observations of the 0.2Jy sample over an area more than four times larger, to fainter optical magnitudes and identifications with O−E≥1.0 in order to be more complete in the range $3.0 < z < 3.5$. We are also studying fainter radio samples ($S > 100mJy$). This will improve the statistics of our sample to lower radio powers and allow a more detailed analysis of the luminosity function and its evolution at $z > 2$.

**Acknowledgments**

IMH acknowledges the receipt of a SERC studentship and RGM thanks the Royal Society for support. We would like to thank Lisa Storrie-Lombardi for reducing the data on GB1745+6227.